\documentclass[conference]{IEEEtran}
\IEEEoverridecommandlockouts
\usepackage{cite}
\usepackage{amsmath,amssymb,amsfonts}
\usepackage{graphicx}
\usepackage{booktabs}
\usepackage{multirow}
\usepackage[english]{babel}

\def\BibTeX{{\rm B\kern-.05em{\sc i\kern-.025em b}\kern-.08em
    T\kern-.1667em\lower.7ex\hbox{E}\kern-.125emX}}
\begin{document}

\title{Error Correction with Systematic RLNC in
Multi-Channel THz Communication Systems}
\author{\IEEEauthorblockN{Cao Vien Phung, Anna Engelmann and Admela Jukan}
\IEEEauthorblockA{Technische Universit\"at Braunschweig, Germany\\
Email: \{c.phung, a.engelmann and a.jukan\}@tu-bs.de
}}
\maketitle

\begin{abstract}
The terahertz (THz) frequency band (0.3-10THz) has the advantage of large available bandwidth and is a candidate to satisfy the ever increasing mobile traffic in wireless communications. However, the THz channels are often absorbed by molecules in the atmosphere, which can decrease the signal quality resulting in high bit error rate of received data. In this paper, we study the usage of systematic random linear network coding (sRLNC) for error correction in generic THz systems with with $2N$ parallel channels, whereby $N$ main high-bitrate channels are used in parallel with $N$ auxiliary channels with lower bit rate. The idea behind this approach is to use coded low-bit rate channels to carry redundant information from high-bit rate channels, and thus compensate for errors in THz transmission. The analytical results evaluate and compare the different scenarios of the THz system in term of the amount of coding redundancy, a code rate, transmission rate of auxiliary channels, the number of THz channels, the modulation format and transmission distance as required system configurations for a fault tolerant THz transmission.
\end{abstract}

\section{Introduction}\label{introduction}
The terahertz (THz) frequency band ranging from $0.3$ THz to $10$ THz has evolved as a prime candidate to satisfy the exponentially growing data volumes in wireless networks. However, the quality of received THz signal, e.g., Signal-to-Noise ratio (SNR), is always a challenge resulting in a high bit error rate (BER) with a increasing signal modulation level and THz transmission distance. Also, increase in BER is often caused by  molecule absorption in the atmosphere \cite{7321055, 7444891}. To this end, the research community gives novel methods to overcome the problem of high BER of THz signal when sending data over longer transmission distances with high modulation format, and under atmospheric effects.

So far, the authors in \cite{Zhou:17, Rout:2016, Mittendorff:2017,Khalid:2016, Ullah:2019, Jiang:2018, Lin:2015, Yan:2019} studied the key components of high-speed THz communication with compensated path attenuation, e.g., fast and efficient amplitude and phase modulators, high gain and massive multiple-input multiple-output (MIMO) antennas and the efficient THz beamforming. In \cite{7444891}, WiFi and THz channel are combined to utilize the distance estimation of the THz receiver and to measure the relative air humidity. The research on an adequate channel coding method is however still in its infancy. Currently,  the state-of-the-art Forward Error Correction (FEC) Codes such as Reed Solomon, Low Density Parity Check (LDPC) and Polar code are under discussion \cite{wehn_norbert_2018_1346686, wehn_norbert_2019_3360520}, and their implementations without reducing the THz channel throughput is an open challenge.

In this paper, we use the idea of deploying generic $2N$ parallel channels to enhancing the overall throughput of THz transmission. We  propose to use sRLNC, a coding technique presented in \cite{5513768}, in addition to a generic low complexity FEC, as proposed by the standard \cite{Wang:2006}, and in parallel we send the coding redundancy generated during encoding process for erasure error correction at the receiver. To this end,  out of  $2N$ channels we deploy $N$ main channels and $N$ auxiliary (coding redundancy) channels. Our system configures  the source data to be transferred over $N$ THz main channels with high bit rate, while the coding redundancy is dispatched over $N$ auxiliary channels with lower bit rate. We analyze and compare the scenarios of THz system with different channel configurations, modulation formats and transmission distances in term of the amount of sRLNC redundancy, the related code rate and the transmission rate of auxiliary channels required for certain reliability of THz transmission. The results show that the main THz channels can use a higher level modulation format and configure the longer transmission distance to achieve a higher overall throughput in the case of auxiliary channels with low bit rate supported.

The remainder of this paper is organized as follows. Sec. \ref{sysdes} designs fault tolerant multi-channel THz transmission system with sRLNC. In Sec. \ref{ana} we analyze the amount of coding redundancy, code rate and bit rate of redundant channels required. Sec. \ref{num} present numerical results. Sec. \ref{conclusion} concludes the paper.
\begin{figure*}[!ht]
\centering
\includegraphics[width=2\columnwidth]{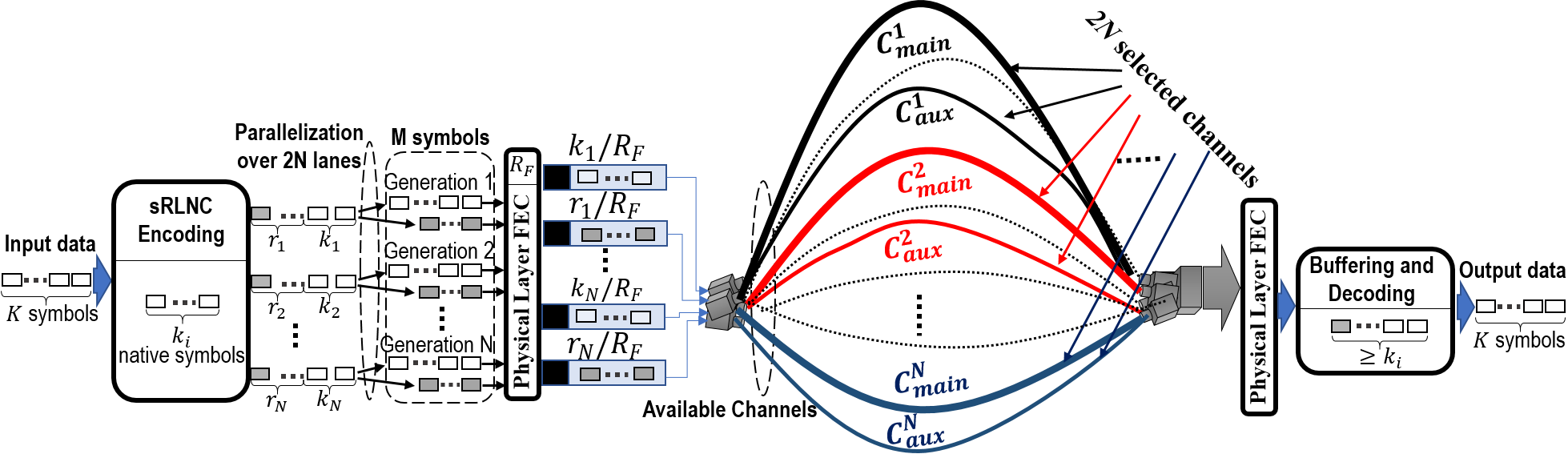}%
  \vspace{-0.1cm}
  \caption{Multi-Channel THz transmission system with systematic random linear network coding (sRLNC).}
  \vspace{-0.6cm}
  \label{codingscenario}
\end{figure*}
\section{System design}\label{sysdes}
In this section, we show a fault tolerant multi-channel THz transmission system in configuration with $2N$ channels and sRLNC under a reference architecture  in Fig. \ref{codingscenario}. This proposal is extended from the 2-channel THz transmission system in our patent \cite{patent}. The notations of the paper  are summarized in Table \ref{tab:table1}.

As shown in Fig. \ref{codingscenario}, source data are distributed and simultaneously transmitted over $2N$ parallel THz channels, where $N$ channels use for original data and $N$ channels are for redundant data. Encoding with sRLNC and traffic  parallelization are performed in the source prior to THz transmission.

At the source, the input native data are a bit stream structured into a sequence of symbols, i.e. $K$ symbols, where each symbol has a fixed size of $s$ bits. All native symbols will be temporarily stored for the encoding process at the encoding buffer. For the encoding process, we define a generation as a set of symbols used to code and decode together at the source and receiver, respectively, where each generation is divided from the input data stored in the encoding buffer. In our example of Fig .\ref{codingscenario}, a generation contains $k$ native symbols encoded together by sRLNC, where the sRLNC is performed over a finite field $\mathbb{F}_{2^s}$, coding coefficients is randomly chosen and generation size is constant. For simplicity, in Fig. \ref{codingscenario} we only show $N$ generations, where any $i^{th}$ generation has $k_i$ native symbols. During encoding process, $r$ additional symbols are generated with linear combinations of a generation with $k$ native symbols along with coding coefficients, however, the native symbols keep unchanged and their size as well as the size of any redundant symbols keeps $s$ bits. Thus, $r_i$ redundant symbols of any $i^{th}$ generation are generated by forming linear combinations from $k_i$ symbols. The number of native symbols of any generation $i$ is fixed, i.e., $k_i=k_j=k$, whereby the amount of redundant traffic depends on the expected BER on its related main channel, i.e., $r_i \neq r_j$.

After sRLNC encoding process, $M=K+R$ symbols, where $K=\sum_{i=1}^Nk_i$ is the total size of generation data and $R=\sum_{i=1}^Nr_i$ is total amount of redundancy, need to be parallelized and distributed over $2N$ channels, i.e., $N$ high bit rate main THz channels and $N$ low bit rate auxiliary channels. The parallelization process includes two steps. In the first step, each generation is distributed over $N$ lanes, while, in the second step, the redundant symbols are separated from the native symbols requiring two lanes per encoded generation and $2N$ lanes in total. After the parallelization of any $N$ generations over $2N$ lanes, FEC per lane is required. In this paper, we do not consider any specific FEC in the physical layer, while sRLNC is applicable in the higher layer (e.g., Ethernet) to complement any FEC mechanisms with code rate $R_F$, when FEC either fails or yields insufficient performance. As a result, before the THz signal transmission, any $k_i$ native symbols and their related redundancy are packetized and protected by FEC with code rate $R_F$, leading to the size of each generation $i$ and its redundancy increased as $\frac{s \cdot k_i}{R_F}$ bits and $\frac{s \cdot r_i}{R_F}$ bits, respectively. 

Finally, any generation $i$ with $k_i$ native symbols is sent on the $i^{th}$ main channel and its related redundancy $r_i$ is transferred on the corresponding $i^{th}$ auxiliary channel. All generations and their redundancy are dispatched at the same time on the transmission channels. For our proposed THz system, the high-level modulation format is used for main channels to achieve a high bit rate, while the related auxiliary channels with lower bit rate are for reliably transferring additional coded symbols to correct error symbols happened on the main channels. Some parameters, such as modulation level, transmission distance, signal power, etc., are necessary to be reasonably tuned on the main and related auxiliary THz channels to address symbol errors occurred during Thz transmission.

At the receiver, the error symbols will be deleted through error detector of FEC mechanism, when these erroneous bits could not be corrected. Therefore, after FEC those lost symbols can be replaced by the sRLNC redundancy from the same coded generation transmitted on the auxiliary channels. The decoding process for recovery of each generation performed by Gauss-Jordan elimination is only successful when at least $k_i$ out of $k_i+r_i$ symbols from any coded generation $i$ arrive at decoder. Finally $N$ decoded generations are serialized. Thus, in order to temporarily store symbols until a generation is complete and decoding process can start, the buffering at the receiver is required. However, it should be noted that the buffering in high speed systems is a big engineering challenge.

\begin{table}[t!]
  \centering
  \caption{List of notations.}
  \label{tab:table1}
  \begin{tabular}{ll}
    \toprule
    Notation & Meaning\\
    \midrule
    %$APL_i$ & Acceptable packet loss rate on the $i^{th}$ main channel.\\
    $P_i^b$ & Residual bit error rate on the $i^{th}$ main channel.\\
    $P_i^s$ & Residual symbol error rate on the $i^{th}$ main channel.\\
    $p_i^e$ & Expected BER on the $i^{th}$ main channel.\\
    $s$ & Symbol size in bits.\\
    $R_F$ & Code rate of FEC code.\\
    $R_L$ & Code rate of sRLNC.\\
    $K$ & Total size of generation data in symbols. \\
    $k_i$ & Size of the $i^{th}$ generation in symbols.\\
    $R$ & Total amount of redundancy in symbols. \\
    $r_i$ & Redundancy of the $i^{th}$ generation in symbols.\\
    $M$ & Total size of coded generation in symbols.\\
    $N$ & Total number of main and auxiliary channels.\\
    $C^i_{main}$ & Transmission rate of the $i^{th}$ main channel.\\
    $C^i_{aux}$ & Transmission rate of the auxiliary $i^{th}$ channel.\\
    $\tau^i_{main}$ & Propagation delay on the $i^{th}$ main channel.\\
    $\tau^i_{aux}$ & Propagation delay on the $i^{th}$ auxiliary channel.\\
    $t^i_{main}$ & Transmission delay on the $i^{th}$ main channel.\\
    $t^i_{aux}$ & Transmission delay on the $i^{th}$ auxiliary channel.\\
    $T^i_{main}$ & Total delay on the $i^{th}$ main channel.\\
    $T^i_{aux}$ & Total delay on the $i^{th}$ auxiliary channel.\\
    $d^i_{main}$ & Transmission distance of the $i^{th}$ main channel.\\
    $d^i_{aux}$ & Transmission distance of the $i^{th}$ auxiliary channel.\\
    %$R_{main}^{T}$ & Transmission rate of the $i^{th}$ main channel.\\
    %$R_{aux}^{T}$ & Transmission rate of the $i^{th}$ auxiliary channel.\\
     $c_p$ & Propagation speed of light, $c_p=3\cdot 10^{8}$m/s.\\
%	$M$ & Modulation level.\\
    \bottomrule
  \end{tabular}
\end{table}

\section{Analysis}\label{ana}
This section gives an analysis of code rate and minimal transmission rate of auxiliary channels in the case of sRLNC applied in the proposed multi-channel THz transmission system with $2N$ channels. Without loss of generality, assume that the minimal Hamming distance of a generic FEC code with a code rate $R_F$ for main channels and auxiliary channels are $\Delta^k_{min}=\frac{k_i \cdot s}{R_F}-(k_i \cdot s)$ and $\Delta^r_{min}=\frac{r_i \cdot s}{R_F}-(r_i \cdot s)$, respectively. We assume bit errors of redundant data on the auxiliary channels can be corrected during FEC operation and their residual BER is equal to $0$. Thus, the Hamming distance can be considered as

\begin{equation}\label{ham}
\Delta_{min} =\frac{k_i \cdot s}{R_F}-(k_i \cdot s) \equiv  \Delta^k_{min}
 \end{equation}

Based on Eq. (\ref{ham}), for each main channel, the number of erroneous bits that can be detected is given

\begin{equation}\label{t_e}
t_e=\Delta_{min}-1
 \end{equation}
, and the number of erroneous bits that can be corrected is expressed

\begin{equation}\label{t_k}
\left\{\begin{matrix}
t_k=\frac{\Delta_{min}-2}{2} & if \; \Delta_{min} \; is \; even \\ 
t_k=\frac{\Delta_{min}-1}{2} & if \; \Delta_{min} \; is \; odd
\end{matrix}\right.
 \end{equation}
, then the expected residual BER $P_i^b$ on any $i^{th}$ main transmission channel after FEC process of correcting bit errors with expected BER $p_e^i$ at receiver can be calculated as

\begin{equation}\label{eqBER}
 P^b_i = \frac{(k_i \cdot s \cdot p_e^{i}) - (R_F \cdot t_k)}{k_i \cdot s}
 \end{equation} 
, where all bit errors of the $i^{th}$ main channel are totally corrected by FEC in the case of $P^b_i \leq  0$. Using Eq. (\ref{eqBER}), the residual symbol error rate on any $i^{th}$ main channel caused by some of bit errors that cannot be corrected by FEC process is determined as

 \begin{equation}
P^s_i=1-(1-P^b_i)^{s}
 \end{equation}
 
As a result, if the expected number of symbol errors per any generation $i$ is $P^s_i \cdot k_i$, then to successfully decode source data, receiver needs at least $P^s_i \cdot k_i$ redundant coded symbols sent over the corresponding $i^{th}$ auxiliary channel, i.e.,

 \begin{equation}\label{redund}
 \forall i \in N:\text{      }   r_i\geq P^s_i\cdot k_i 
  \end{equation} 
, where the number of additional coded symbols $r_i$ sent on the $i^{th}$ auxiliary channel is considered as a function of the residual symbol error rate of the $i^{th}$ main channel, while any generation size $k_i$ is constant.

Thus, the total size of native data symbols sent over all main THz channels can be calculated as.
 \begin{equation}\label{gen}
K=\sum_{i=1}^{N}k_i  
\end{equation}
and the total number of redundant symbols sent over all additional auxiliary channels is determined with Eq. \eqref{redund} as.
 \begin{equation} \label{redundantfomula}
R=\sum_{i=1}^{N}r_i =\sum_{i=1}^{N}(P^s_i \cdot k_i)
\end{equation}
, where $R \geq 0$. Thus, the total number of the native and redundant symbols sent over all $2N$ THz channels can be determined as   
\begin{equation}\label{total}
 M=K+R=\sum_{i=1}^{N}(k_i+r_i) 
 \end{equation}
, where $k_i \leqslant  k_i+r_i$, $r_i\geq 0$ and $K \leqslant M$. Note, if $N=1$, there are in total $2$ THz channels utilized, where the total number of redundant symbols transferred on the auxiliary channel is $R = r_1=P^s_1 \cdot k_1=P^s_1 \cdot K$. While the code rate $R_F$ of the generic FEC is constant, the overall code rate $R_L$ of sRLNC is function of channels' configuration and the number of additional symbols sent to protect all $N$ generations.
\begin{equation}\label{CR}
R_L=\sum_{i=1}^{N}\frac{k_i}{k_i+r_i}=\frac{K}{M}
 \end{equation}

For any main channel $i$, $k_i$ native symbols of generation $i$ arriving at the receiver after time interval are \begin{equation}\label{Tmain}
T_{main}^i=t_{main}^i+\tau_{main}^i=\frac{k_i \cdot s}{R_F \cdot C_{main}^i} + \frac{d_{main}^i}{c_p}
 \end{equation}
, where $t_{main}^i$ is the transmission delay of $k_i$ native symbols of generation $i$, $\tau_{main}^i$ stands for the propagation delay of the channel $i$, $C_{main}^i$ denotes a transmission rate of the main channel $i$, $d_{main}^i$ is the transmission distance between sending and receiving THz antenna of the main channel $i$ and $c_p$ is the propagation speed of light. 

For any auxiliary channel $i$, $r_i$ redundant coded symbols of generation $i$ arriving at the receiver after time interval are \begin{equation}\label{Taux}
T_{aux}^i=t_{aux}^i+\tau_{aux}^i=\frac{r_i \cdot s}{R_F \cdot C_{aux}^i} + \frac{d_{aux}^i}{c_p}
 \end{equation} , where $t_{aux}^i$ is the transmission delay of $r_i$ additional symbols of generation $i$, $\tau_{aux}^i$ represents the propagation delay of auxiliary channel $i$, $d_{aux}^i$ is the transmission distance between sending and receiving THz antenna of the auxiliary channel $i$ and $C_{aux}^i$ stands for a transmission rate of auxiliary channel $i$.
 
The THz system should be configured so that the arriving time interval of $k_i$ native symbols at the receiver with Eq. (\ref{Tmain}) is equal to that of $r_i$ additional symbols with Eq. (\ref{Taux}) to avoid a very large buffer, i.e., $T^i_{main}=T^i_{aux}$, or 
\begin{equation}\label{TmainTaux}
\frac{k_i \cdot s}{R_F \cdot C_{main}^i} + \frac{d_{main}^i}{c_p}=\frac{r_i \cdot s}{R_F \cdot C_{aux}^i} + \frac{d_{aux}^i}{c_p}
 \end{equation}. Therefore, the transmission rate of any auxiliary channel $i$ is a function of redundancy and transmission distance of main and auxiliary channels and can be determined from Eq. \eqref{TmainTaux} as
 \begin{equation}\label{rateaux}
C_{aux}^{i}=\frac{r_i \cdot s\cdot c_p\cdot C_{main}^i}{R_F \cdot C_{main}^i \cdot (d_{main}^i-d_{aux}^i)+c_p \cdot k_i \cdot s}
 \end{equation}

Generally, $C_{aux}^i \geq 0$, whereby $C_{aux}^i=0$ means that it is unnecessary to create an auxiliary channel $i$ for additional symbols. Based on Eq. \eqref{rateaux}, we can derive a constraint $R_F \cdot C_{main}^i(d_{main}^i-d_{aux}^i)+c_p \cdot k_i \cdot s>0$, then the transmission distance of any $i^{th}$ auxiliary channel has to be limited as 
 \begin{equation}\label{dlimit}
d_{aux}^i < \frac{k_i \cdot s \cdot c_p}{R_F \cdot C_{main}^i} + d_{main}^i
 \end{equation}
The purpose for the transmission distance limit of auxiliary channels is to maximally reduce the buffer size at the receiver.

Assume that we configure the transmission rate of all main channels is the same, i.e., $C_{main}^i=C_{main}^j=C_{main}$, all generations have the same size, i.e. $k_i=k_j=k$, the transmission distance of all main channels are the same, i.e., $d_{main}^i=d_{main}^j=d_{main}$ and the transmission distance of all auxiliary channels are the same, i.e., $d_{aux}^i=d_{aux}^j=d_{aux}$. Then, based on Eq. \eqref{rateaux}, the total transmission rate $C_{aux}^t$ of all auxiliary channels is given by  \begin{equation}\label{totalauxrate}
C_{aux}^t = \sum_{i=1}^N C^i_{aux}=\frac{s\cdot c_p\cdot C_{main} \cdot\sum_{i=1}^N r_i}{R_F \cdot C_{main} \cdot (d_{main}-d_{aux})+c_p \cdot k \cdot s}
 \end{equation}
\section{Numerical results}\label{num}
In this section, we analyze the results of THz system scenarios with $2N=2$ channels and $2N=4$ channels for the code rate of sRLNC $R_L$ and for the transmission rate of auxiliary channels calculated with Eqs. \eqref{CR}, \eqref{redund}, \eqref{eqBER} and \eqref{rateaux}, \eqref{totalauxrate}, respectively, whereby two channel types, i.e.,  Channel B ($660-695$ GHz) and channel C ($855-890$ GHz), are considered and simulated by the raised cosine filter with roll-off factor $0.4$ in \cite{7444891}. The transmission rate of any main channel $i$ used is $C_{main}^i=C_{main}=25$ GBd/s$=2 \cdot 10^{11} \cdot L$ bps, where $L$ denotes the number of bits per symbol for a certain modulation level. The expected BER $p_e^i$ of any main channel $i$ collected from \cite{7444891} is a function of THz channel type, distance and modulation format. Assume that any auxiliary channel $i$  in our evaluation is configured in term of transmission distance, modulation format, channel type, etc., so that the residual BER and co-channel interference on that one is approximately equal to $0$. The symbol size is set to $s=8$ bits. The total number of symbols is set to $K=100$ symbols. The code rate $R_F=0.73$ of generic FEC is used for all channels. The transmission distance of main channels between sending and receiving antennas is set up in interval $[200,2000] cm$ and the modulation formats investigated are 16PSK, 8PSK, QPSK and BPSK. The transmission distance of all main channels (B and C channel) as well as all auxiliary channels is configured the same, i.e., $d_{main}^i=d_{main}^j=d_{main}$ and $d_{aux}^i=d_{aux}^j=d_{aux}$, respectively. All of main channels use the same modulation format.

For a THz system configuration of $2N=2$ channels, the Hamming distance calculated with Eq. \eqref{ham} is $\Delta_{min}=\frac{100 \cdot 8}{0.73}-(100 \cdot 8)=296$ bits for any generation of $K=k_i=100$ symbols. Then, the number of erroneous bits that can be corrected by FEC and calculated with Eq. \eqref{t_k} is $t_k=\frac{296-2}{2}=147$ bits.

For a THz system configuration of $2N=4$ channels, the Hamming distance of any channel $i$ defined in Eq. \eqref{ham} is set to $\Delta_{min}=\frac{50 \cdot 8}{0.73}-(50 \cdot 8)=148$ bits, where $k_i=k=50$, and the number of erroneous bits on each of two main channels that can be corrected by FEC and calculated with  Eq. \eqref{t_k} is $t_k=\frac{148-2}{2}=73$, i.e., $146$ bits in total. 

\begin{figure}[t]
\centering
\includegraphics[width=1\columnwidth]{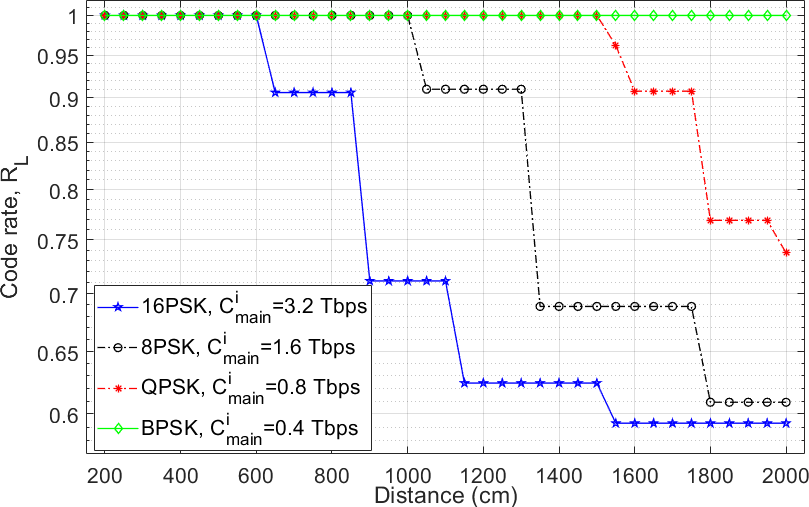}
\caption{Code rate vs. transmission distance $d_{main} $ in configuration with two main channels (B and C), i.e., $N=2$.}
\label{4channels}
\vspace{-0.3cm}
\end{figure}

Figure \ref{4channels} presents the code rate for the fault tolerant THz transmission system with $2N=4$ channels. Thus, there are two main channels (B and C channel) and any two auxiliary channels with lower transmission rate. For BPSK, no redundancy is required and the code rate is always $R_L=1$ because all bit errors on the two main channels (B and C) can be corrected by FEC. In the case of QPSK, the additional redundancy needs to support the main channels, when transmission distance $d_{main} \geq 1550cm$. That results in the code rate, which varies from $1$ to $0.74$. The minimal code rate of $0.61$ in case of 8PSK is reached when transmission distance $d_{main}$ is larger than or equal to $1800cm$. With 16PSK, the code rate changes from $1$ to $0.59$, when transmission distance $d_{main}$ increases from $600cm$ to $1550cm$. We observe that with an increasing distance $d_{main}$ and an increasing modulation level the code rate $R_L$ reduces. When we use the modulation format on the two main channels, the code rate $R_L$ can be independent and achieved a maximal value, i.e., $R_L \approx 1$. In addition, it is possible to use modulation format up to 16PSK with a code rate $R_L \leq 0.91$.

\begin{figure}[t]
\centering
\includegraphics[width=1\columnwidth]{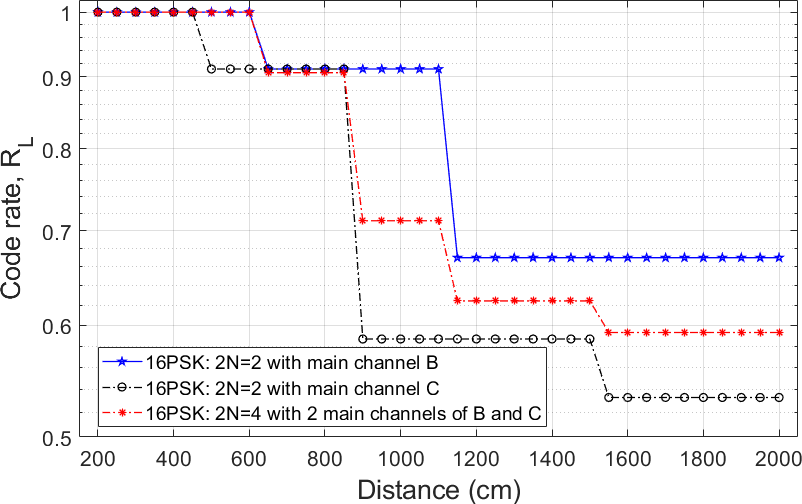}
\caption{Comparison of code rates as a function of transmission distance $d_{main} $ in configuration with one main channel B; one main channel C and two main channels (B and C), i.e., $N=2$.}
\label{comparison}
\vspace{-0.3cm}
\end{figure}

Figure \ref{comparison} shows a comparison of the code rates in the scenarios of $2N=2$ channels with main channel B, $2N=2$ channels with main channel C and $2N=4$ channels with the two main channels B and C. The presented results are for the modulation format 16PSK only, which enable to achieve a high transmission rate. The highest and lowest code rate can be considered for the case of $2N=2$ channels with C main channel and B main channel, respectively. The configuration with $2N=4$ channels, where two main channels B and C are utilized simultaneously, provides a trade off. Generally, using channel B as a main channel with 16PSK and a code rate of $0.9$, the transmission distance can reach up to $d_{main}=1100 cm$ in configuration with $2N=2$ channels.

\begin{figure}[t]
\centering
\includegraphics[width=1\columnwidth]{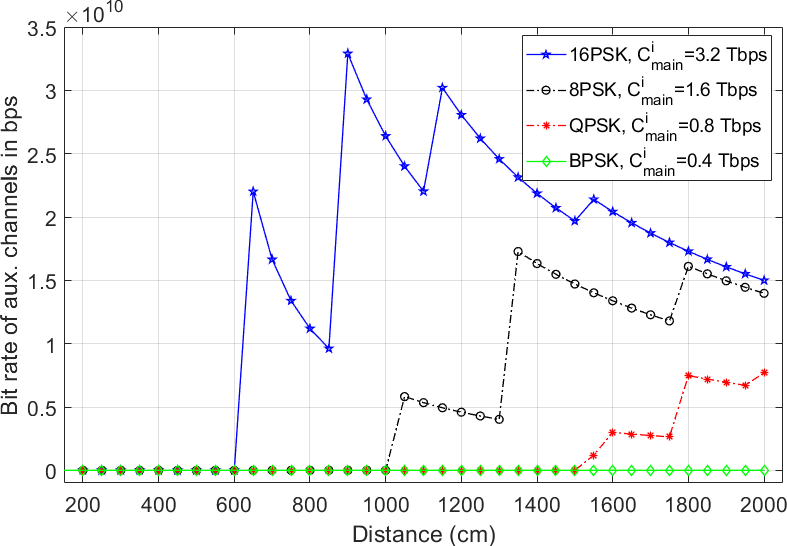}
\caption{Total transmission rate of all auxiliary channels in configuration with $N=2$ vs. transmission distance $d_{main}$ of main channels (B and C), where $d^1_{main}=d^2_{main}$ and $d^1_{aux}=d^2_{aux}=500 cm$.}
\label{BitrateNchannels}
\vspace{-0.3cm}
\end{figure}

Next, we evaluate the transmission rate of auxiliary channels required for a reliable THz system with configuration of $2N=4$ channels in Fig. \ref{BitrateNchannels}, 
whereby channel B and channel C are utilized as main channels. The transmission rate of any auxiliary channel $i$ calculated with Eq. \eqref{rateaux} is a function of a number of redundant symbols $r_i$ and a transmission distance of the main and auxiliary channels, $d^i_{main}$ and $d^i_{aux}$. 
For simplicity, we show the total transmission rate $C_{aux}^t$ over all auxiliary channels using  Eq. \eqref{totalauxrate}. The mean transmission rate of an auxiliary channel can be calculated as $C_{aux}^i=\frac{C_{aux}^t}{N}$, where $N$ is the number of auxiliary channels utilized. The transmission distance of any auxiliary channel $i$ is fixed as $d_{aux}^i=d_{aux}=500cm=5m$. Next, we investigate the total transmission rate $C^t_{aux}$ considering different transmission distances and modulation formats of main THz channels. 
For BPSK, $C^t_{aux}=0$ for all transmission distances of main THz channels, therefore we do not need to use any auxiliary channels because FEC code can correct all error bits from all main THz channels. 
Observe Fig .\ref{BitrateNchannels}, we see that the behavior of 16PSK, 8PSK and QPSK is quite similar, whereby 16PSK, 8PSK and QPSK have a fluctuation in interval $[600,2000]cm$, [$1000,2000cm$] and [$1500,2000cm$], respectively. 
We take an example of 16PSK to explain the behavior. $C_{aux}^t$ gradually decreases from $2.201 \cdot 10^{10}$ bps to $0.962 \cdot 10^{10}$ bps when the transmission distance $d_{main}$ increases because the total amount of redundancy $R$ sent, which is constant, i.e., $11$ symbols, in transmission distance interval $d_{main}$ $[650,850]cm$ and also at 
the same time the increasing transmission distance $d_{main}$ of any main channel $i$ leads to a decreasing total transmission rate $C_{aux}^t$ of all auxiliary channels, which have a fixed transmission distance $d_{aux}^i=d_{aux}=5m << d_{main}^i=d_{main}$, and then achieves a peak at $3.290 \cdot 10^{10}$ bps because the increasing number of redundant symbols required, i.e., from $11$ additional symbols at $d_{main}= 850cm$ to $41$ additional symbols at $d_{main}=900cm$. After the transmission distance of $d_{main}=900cm$, the behavior can be similarly explained as in interval $[650,900]cm$. 

Fig. \ref{BitrateNchannels} shows that the total transmission rate $C_{aux}^t$ of auxiliary channels increases when the modulation level on the main THz channels increases. Additionally, the results confirm our statement in Eq. \eqref{dlimit} that  when $d_{aux}<<d_{main}$, the  total transmission rate $C_{aux}^t$ of auxiliary channels can reach the minimal value, whereby the symbol data of the main channels and the auxiliary channels will simultaneously arrive to maximally reduce the buffering overhead at receiver.

%\begin{figure}[t]
%\centering
%\includegraphics[width=1\columnwidth]{figures/BitrateNchannelsdauxequaldmain}
%\caption{Total transmission rate of all auxiliary channels vs. transmission distance of two main channels B and C, when $d^1_{aux}=d^2_{aux}=d_{main}^1=d_{main}^2$.}
%\label{BitrateNchannelsdauxequaldmain}
%\vspace{-0.1cm}
%\end{figure}

%\begin{figure}[t]
%\centering
%\includegraphics[width=1\columnwidth]{figures/comparisonBitrate}
%\caption{Total transmission rate of all auxiliary channels vs. transmission distance of two main channels B and C, when $d^1_{aux}=d^2_{aux}=d_{main}^1=d_{main}^2$.}
%\label{comparisonBitrate}
%\vspace{-0.1cm}
%\end{figure}

\begin{figure}[t]
\centering
\includegraphics[width=1\columnwidth]{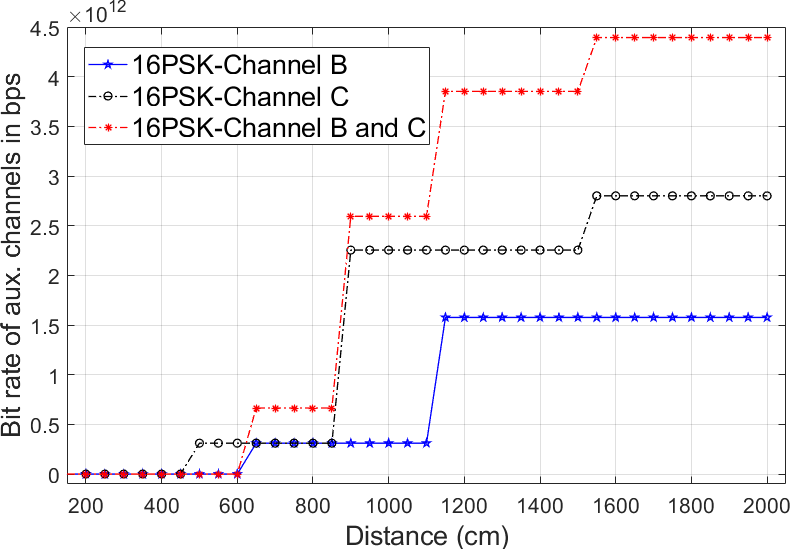}
\caption{Total transmission rate of all auxiliary channels in configuration with one B main channel, one C main channel and with $N=2$ (B and C channels) vs. transmission distance $d_{main}$ of main channels, where $d^1_{aux}=d^2_{aux}=d_{main}^1=d_{main}^2$.}
\label{comparisonBitratedmainequaldaux}
\vspace{-0.3cm}
\end{figure}

Next, we consider and compare the total transmission rate of auxiliary channels for the THz transmission system with $2N=2$ channels configured main channel either as channel B or as channel C and with $2N=4$ channels configured two main channels including channel B and channel C, when the transmission distance of main and auxiliary channels are the same, i.e., $d_{main}=d_{aux}$. For analyzed results with high transmission rate of the main THz channel, Fig. \ref{comparisonBitratedmainequaldaux} only investigate the modulation format of 16PSK and is a function of the transmission rate $d_{main}$ and main channel configuration. Based on Eq. \eqref{totalauxrate}, we see that the total transmission rate $C_{aux}^t$ of auxiliary channel only depends on the number of additional symbols when $d_{main}=d_{aux}$. As BER increases, the increasing number of coding redundancy needs a high transmission rate of auxiliary channels. Also, if the transmission distance of main channels $d_{main}$ insignificantly increases, then the number of additional symbols required as well as the total transmission rate of auxiliary channels $C_{aux}^t$ will keep unchanged. For instance of the scenario with $2N=4$ channels, the number of redundant symbols keeps unchanged with $11$ symbols in interval $[650,850] cm$ requiring the constant total transmission rate $C_{aux}^t=0.665 \cdot 10^{12}$ bps. We observe that the transmission rate of auxiliary channel in the THz system $2N=2$ channels with the main channel configured as channel C is always higher than that in the case of the THz system $2N=2$ channels with main channel B because  the symbols sent on the channel C often occur  more error bits requiring more additional coded symbols for erroneous symbol recovery at receiver. On the other hand, although the configuration with $2N=4$ channels provides a trade off, its total transmission rate is the highest because its generation size with $k=50$ symbols is less than $K=100$ symbols of the system with $2N=2$ channels as derived from Eq. \eqref{totalauxrate}. For example, with the same transmission distance of $d_{main}=d_{aux}=2000cm$, the THz system of $2N=2$ channels with main channel B, $2N=2$ channels with main channel C and $2N=4$ channels need auxiliary channels with the total transmission rate $C_{aux}^t=1.577 \cdot 10^{12}$ bps, $2.800 \cdot 10^{12}$ bps and up to $4.394 \cdot 10^{12}$ bps.

\section{Conclusion}\label{conclusion}
In this paper, we propose to use $2N$ channels with systematic RLNC to improve the error correction in THz communications. Our idea is to complement a generic low complexity FEC code by a low complexity sRLNC, whereby multiple THz channels can be simultaneously operated. For our design, we send the native data over $N$ main channels and coding redundancy over $N$ auxiliary channels in parallel. There are other wireless or wireline
technologies that can use for auxiliary channels, such as the Radio frequency (RF) or Free Space Optics (FSO) required in Mbps area, but these ones also suffer from atmospheric, such as fog, rain, etc, or even re-using THz frequency band. The auxiliary channels should be configured the distance and modulation type and channel type so that their residual BER and co-channel interference approximately achieve $0$ for a certain reliability. Our analysis evaluates and compares the scenarios of the THz system with the different number of channels in term of code rate of sRLNC, coding redundancy and the transmission rate of auxiliary channels required for a certain reliability transmission. The results show that the main channels can use a high-level modulation format and send data over longer distances, when the auxiliary channels can send additional symbols with lower transmission rate for the erasure error correction at the receiver.

\section*{Acknowledgment}
This work was partially supported by the DFG Project Nr. JU2757/12-1, "Meteracom: Metrology for parallel THz communication channels."

\bibliographystyle{IEEEtran}
\bibliography{nc-rest}

\end{document}